\documentclass[twocolumn]{article}

\usepackage{amsmath}
\usepackage{amssymb}

\usepackage{cite}

\usepackage[T1]{fontenc}
\usepackage{newtxtext,newtxmath}
\usepackage{microtype}
\usepackage{framed,multirow}

\usepackage{latexsym}
\usepackage{url}
\usepackage{xcolor}
\usepackage{geometry}
\makeatletter
\AtBeginDocument{%
  \parindent=3.8mm%
  \renewcommand\normalsize{%
    \@setfontsize\normalsize{9}{10}%
    \abovedisplayskip 6\p@ \@plus1\p@ \@minus1\p@
    \abovedisplayshortskip 3\p@ \@plus1\p@
    \belowdisplayshortskip 3\p@ \@plus1\p@
    \belowdisplayskip \abovedisplayskip
    \let\@listi\@listI
  }%
  \normalsize
}
\makeatother

\usepackage{titlesec}
\titleformat{\section}
  {\normalfont\normalsize\bfseries}
  {\thesection}{0.6em}{}

\titleformat{\subsection}
  {\normalfont\normalsize\mdseries\itshape}
  {\thesubsection}{0.6em}{}

\titleformat{\subsubsection}
  {\normalfont\footnotesize\mdseries\itshape}
  {\thesubsubsection}{0.6em}{}

\usepackage{etoolbox}
\AtBeginEnvironment{thebibliography}{%
  \providecommand{\url}[1]{}%
  \providecommand{\href}[2]{#2}%
}

\usepackage{physics,graphicx,bm,svg}
\usepackage{algorithm,algorithmic}
\usepackage{printlen}
\usepackage{authblk}

\newcommand{\mat}[1]{\boldsymbol{#1}}
\newcommand{\vect}[1]{\boldsymbol{#1}}
\newcommand{\isreal}{\in \mathbb{R}}
\newcommand{\covar}[1]{\mat{\Sigma_{#1}}}
\newcommand{\precis}[1]{\mat{\Sigma^{-1}_{#1}}}

\DeclareRobustCommand{\bigO}{%
  \text{\usefont{OMS}{cmsy}{m}{n}O}%
}

\usepackage{hyperref}

\title{Bayesian Inference of Flame Impulse Responses}%

\author[1]{Matthew Yoko}
\author[2]{Wolfgang Polifke}
\affil[1]{Department of Engineering, University of Cambridge, Cambridge, CB2 1PZ, United Kingdom.}
\affil[2]{Department of Engineering Physics and Computation, Technical University of Munich, Boltzmannstr. 15, Garching, 85748, Germany.}
\date{} 

\begin{document}
\makeatletter
\twocolumn[%
\begin{@twocolumnfalse}
\maketitle

\begin{abstract}

The impulse response of a flame to acoustic velocity perturbations is a key quantity for predicting thermoacoustic stability, but its identification from sparse, noisy observations requires solving an ill-posed inverse convolution problem. This is typically achieved with system identification methods, which require hand-tuning of regularization, model order, and sampling parameters, and provide no principled mechanism for incorporating prior physical knowledge. In this paper, we reformulate the identification problem within a Bayesian framework. The impulse response is represented as a physically motivated distributed time delay model, whose parameters correspond to convective delays and dispersive broadening. For a given number of pulses, the model parameters are inferred from the data using Bayesian parameter inference. The number of pulses is then selected using Bayesian model comparison, which balances data fit against model complexity to identify the simplest model capable of explaining the data. The framework is demonstrated on broadband-forced large eddy simulation data from a turbulent swirl-stabilized burner. Bayesian model comparison selects a three-Gaussian impulse response for this flame, consistent with physical interpretations in previous work. Compared with system identification, the Bayesian approach produces impulse responses with fewer spurious features and enables straightforward enforcement of a known low-frequency gain. Finally, we show that the Bayesian approach is robust to significant reductions in recording length, making it appealing for impulse response identification from costly simulations, where there is an incentive to minimize computational cost.
    
\end{abstract}

\vspace{1em} 
\end{@twocolumnfalse}
]
\makeatother

\section{Introduction}
\label{sec:introduction}

The linear response of a premixed flame to an arbitrary acoustic velocity disturbance is entirely described by the flame's impulse response, or equivalently its transfer function \cite{Polifke2020,Schuller2020a}. For a given perturbation in velocity $u'$, the resulting perturbation in heat release rate $q'$ can be expressed as the convolution of $u'$ with the impulse response $h$:
\begin{equation}
    q' = u' * h = \int_{-\infty}^{\infty} u'(t-\tau)h(\tau)\mathrm{d}\tau
    \label{eq:continuous-conv}
\end{equation}

Given a flame's impulse response and an acoustic model of the system, it is therefore possible to predict the thermoacoustic stability of a system using linear analysis. The impulse response, however, is generally unknown and not directly measurable. A core challenge in thermoacoustics is to determine the impulse response or transfer function of a flame from sparse, noisy observations of $u'$ and $q'$, either from experiments or from numerical simulations. This requires solving an inverse convolution problem:
\begin{equation}
    \text{given } (u',q'), \text{ find } h \text{ such that } q' = u' * h.
    \label{eq:inverse-conv}
\end{equation}

Inverse convolution problems are generally ill-posed, meaning that small errors in the data can lead to large errors in the estimated impulse response. Therefore, care must be taken in how the problem is formulated and solved.

A common approach is to excite the system with a broadband signal, such as filtered white noise, and record the time-domain response of the heat release rate \cite{Polifke2001,Polifke2014}. By capturing $M$ discrete samples of the input-output data $(u_m,q_m)$ at uniform intervals $\Delta t$, the discrete convolution with an impulse response of length $L$ can be expressed as:
\begin{equation}
    q_m = \sum_{l=0}^{L-1} u_{m-l} h_{l}
    \label{eq:discrete-conv}
\end{equation}
where $q_m = q'(m\Delta t)$, $u_{m-l} = u'((m-l)\Delta t)$, and $h_l = h(l\Delta t)$. Note that we expect the output to only depend on past inputs, i.e. the flame cannot preempt a velocity perturbation. This requires special treatment for the entries of $u_{m-l}$ for which $m-l < 0$. We discuss this in detail in Section \ref{sec:robust}.

The discrete convolution in eq. \eqref{eq:discrete-conv} can be expressed in matrix form as:
\begin{equation}
    \vect{q} = \mat{U}\vect{h}
    \label{eq:linear-inverse}
\end{equation}
where $\vect{q} = [q_0, q_1, \dots, q_{M-1}]^\top$, $\vect{h} = [h_0, h_1, \dots, h_{L-1}]^\top$, and $\mat{U} \isreal^{M \times L}$ is the Toeplitz matrix formed from the discrete input signal $\vect{u} = [u_0, u_1, \dots, u_{M-1}]^\top$. The impulse response can then be estimated by solving the regularized least squares problem:
\begin{equation}
    \hat{\vect{h}} = (\mat{U}^\top \mat{U} + \lambda \mat{I})^{-1} \mat{U}^\top \vect{q}
    \label{eq:least-squares}
\end{equation}  
where $\mat{I}$ is the identity matrix and $\lambda$ is the regularization parameter, which must be hand-tuned to provide sufficient smoothing without excessively biasing the result.

After fitting $\hat{\vect{h}}$, the residual $\vect{e}=\vect{q}-\mat{U}\hat{\vect{h}}$ can be analysed to obtain approximate uncertainty bounds on the impulse response, which can then be propagated to the transfer function \cite{Radack2025a}.

While variants of this approach have been widely used \cite{Radack2025a,Giauque2008,Komarek2010,Tay-Wo-Chong2013,Eder2023a,Wang2025,Ke2024,Bothien2019}, it has several drawbacks. The matrix $\mat{U}^\top \mat{U}$ can be ill-conditioned depending on the properties of the input signal, leading to numerical instability in the inversion, and amplification of small errors. While regularization can ease this, the choice of regularization parameter $\lambda$ is arbitrary and can bias the estimated impulse response. Most importantly, this method requires that all information about the impulse response is provided by the data, with no way to incorporate prior knowledge about the physics of the system.

In general, we have some broad but useful prior knowledge about the expected form of the impulse response. For example, we know that the impulse response must be smooth, and that it typically consists of a few characteristic time delays related to the physical processes in the flame \cite{Polifke2020}. We know that these time delays are typically governed by convective transport processes, providing a natural timescale for the impulse response. We know that the impulse response must be causal (i.e. all time delays must be positive), and we know that its integral describes the steady-state gain of the flame, which can often be prescribed based on physical arguments \cite{Polifke2007}.

In this work, we reformulate this inverse convolution problem as a Bayesian inference problem, allowing us to combine the observed data with prior knowledge about the expected form of the impulse response. By proposing a physically-motivated parametric model for the impulse response, such as a distributed time delay model \cite{Polifke2020}, we can encode physical constraints such as smoothness, causality and characteristic timescales. This leads to a more robust and interpretable estimation procedure that is less sensitive to noise and numerical instability.

This paper is structured as follows. In Section \ref{sec:model}, we introduce the physically motivated model for the impulse response. In Section \ref{sec:bayesian}, we formulate the inverse problem in a Bayesian framework, defining the prior, likelihood, and posterior distributions. In Section \ref{sec:robust}, we describe a robust solution procedure for the Bayesian inverse problem. In Section \ref{sec:results}, we present results from applying this framework to large eddy simulation (LES) data, before concluding in Section \ref{sec:conclusions}.

\section{Physically motivated model of the impulse response}
\label{sec:model}

We have prior knowledge that flame dynamics are typically governed by convective processes with a few characteristic time delays, and often some diffusion or dispersion leading to these delays being distributed in time \cite{Polifke2020}. Several studies have proposed modelling the flame impulse response as a sum of Gaussian delays to capture this behaviour \cite{Schuermans2004,Komarek2010,Bade2013,Polifke2020,Aesoy2020,Blonde2023,Moon2024}, so we adopt this as the parametric model for the impulse response:
\begin{align}
    h(t;N,\vect{a}) &= \sum_{i=1}^N n_i\,(2\pi\sigma_i^2)^{-1/2}\exp\!\left[-\frac{(t-\tau_i)^2}{2\sigma_i^2}\right]
\end{align}
where $N$ is the number of Gaussian pulses in the model and $\vect{a}$ is the vector of model parameters, which comprise: $n_i$, the amplitude, $\tau_i$, the time delay, and $\sigma_i$, the width of each Gaussian pulse. 

We will refer to this as the $N$-$n$-$\tau$-$\sigma$ model of the impulse response. By exploiting this prior knowledge of the physics, the problem stated in eq. \eqref{eq:inverse-conv} has reduced to:
\begin{equation}
    \text{given } (u',q'), \text{ find } \{N, \vect{a}\} \text{ such that } q' = u' * h(t;N,\vect{a})
    \label{eq:parameter-estimation}
\end{equation}
We have therefore reduced the inverse convolution problem, which requires estimating an arbitrary function $h$ in a high-dimensional function space, into a low-dimensional parameter estimation problem. In this parameter estimation problem, we have one discrete parameter $N$ (the number of Gaussian delays) and $P = 3N$ continuous parameters (the amplitudes, time delays, and widths of each Gaussian), each of which must be estimated from the data.

If we have prior knowledge about the low-frequency limit of the flame transfer function, which is often the case \cite{Polifke2007}, the $N$-$n$-$\tau$-$\sigma$ model can be easily modified to enforce this constraint. The low-frequency limit of the transfer function is given by the time-integral of the impulse response, which for the $N$-$n$-$\tau$-$\sigma$ model is simply the sum of the amplitudes. If the low-frequency gain $G$ is known, we can enforce the constraint $\sum_{i=1}^N n_i = G$ by inferring $N$-1 of the amplitudes and setting the final amplitude to $n_N = G - \sum_{i=1}^{N-1} n_i$.

\section{A probabilistic formulation of the inverse problem}
\label{sec:bayesian}

We now formulate the inverse problem in eq. \eqref{eq:parameter-estimation} in a Bayesian framework, allowing us to incorporate further prior information about the parameters themselves. We treat the discrete and continuous parameters separately. 

The $N$-$n$-$\tau$-$\sigma$ model defines a \emph{family} of models indexed by the model order $N$. For a model of a given order, we use Bayesian parameter inference to estimate the $P$ continuous parameters, given the data. We repeat this for all plausible model orders and use Bayesian model comparison to select the most likely model order, given the data. These two steps are now described in detail.

\subsection{Bayesian parameter inference}

\subsubsection{The prior}
\label{sec:prior}

For a model of a given order $N$, we define prior probability density functions (pdfs) over the possible values of each of the continuous parameters $\vect{a} = \{\vect{n},\vect{\tau},\vect{\sigma}\}$. The priors are designed to enforce physical constraints such as positivity of $\vect{\tau}$ and $\vect{\sigma}$, as well as the expected scales of these parameters based on the convective timescales. The priors will be described in detail in a later section, but for now we note that they are chosen to be Gaussian pdfs:
\begin{equation}
    p(\vect{a}|N) = \mathcal{N}(\vect{a};\vect{\mu_a}, \covar{a})
\end{equation}
where $\vect{\mu_a} = [\vect{\mu_n}, \vect{\mu_\tau}, \vect{\mu_\sigma}]$ is the prior mean and $\covar{a}$ is the prior covariance matrix, whose diagonal entries represent the prior uncertainty in each parameter and off-diagonal entries represent correlations between parameters. We typically do not have strong prior knowledge about how the parameters are correlated, so we set the off-diagonal entries to zero.

In general, we know little about the parameter values \textit{a-priori}, but we are typically able to bound them within reasonable ranges based on physical arguments. If we are able to estimate a minimum and maximum expected value for each parameter, this is sufficient to define a Gaussian prior with a mean at the center of the range and a standard deviation such that 99.7\% of the probability mass lies within the range (i.e., plus/minus three standard deviations). This leads to broad but useful priors that can be applied to a wide range of flames without requiring detailed prior knowledge. This will be explored in detail in Section \ref{sec:robust}.

\subsubsection{The likelihood}
\label{sec:likelihood}

To construct the likelihood function, we must define a noise model describing the relationship between the observed data, $\vect{u}$ and $\vect{q}$, and the model predictions $\hat{\vect{q}}(\vect{a}) = \mat{U}\,\vect{h}(\vect{a})$. In the context of forced incompressible LES, the input record $\vect{u}$
is evaluated as the normalized area-integral of the volume flux through a reference plane. This process suppresses turbulent fluctuations, which are the dominant source of noise in the input signal, so we assume that $\vect{u}$ is noise-free. The heat release rate $\vect{q}$, by contrast, is subject to noise and we therefore model the relationship between the observations and the model predictions as:
\begin{equation}
    \vect{q} = \hat{\vect{q}}(\vect{a}) + \vect{\varepsilon},
    \label{eq:obs-model}
\end{equation}
where $\vect{\varepsilon}$ collects all sources of discrepancy between the observed heat release rate and the simplified impulse response model. In experiments, this term is dominated by sensor and acquisition noise. In LES, $\vect{\varepsilon}$ should instead be interpreted as an effective model discrepancy term, comprising (i) the structural error due to the restricted $N$-$n$-$\tau$-$\sigma$ family, (ii) departures from the linear time-invariant assumption (e.g.\ weak nonlinearity or slow time variance), (iii) turbulent fluctuations, and (iv) numerical and post-processing artefacts.

We assume this discrepancy is approximately Gaussian with zero mean, $\vect{\varepsilon} \sim \mathcal{N}(\vect{0}, \covar{q})$, which can be justified both as the aggregate of many small contributions and as the maximum-entropy choice given only a mean and covariance \cite[§19.2.1]{Jaynes2003}.

Under this assumption the likelihood is
\begin{equation}
    p(\vect{q}|\vect{a},N) = \mathcal{N}(\vect{q}; \hat{\vect{q}}(\vect{a}), \covar{q}).
    \label{eq:likelihood1}
\end{equation}

In the simplest case we take $\covar{q}=\sigma_q^2 \mat{I}$, corresponding to temporally uncorrelated discrepancy, where $\sigma_q$ is estimated from the data as described in Section~\ref{sec:NoiseEstimate}. In general, the discrepancy may be temporally correlated, in which case a parametric model for $\covar{q}$ can be introduced, but this is left for future work.

We now expand the likelihood function in order to gain a deeper understanding of its properties:
\begin{equation}
    p(\vect{q}|\vect{a},N) = Z \exp\!\left[-\frac{1}{2}(\vect{q} - \hat{\vect{q}}(\vect{a}))^\top \precis{q} (\vect{q} - \hat{\vect{q}}(\vect{a}))\right]
    \label{eq:likelihood}
\end{equation}
where $Z = |2\pi\covar{q}|^{-1/2}$ is the normalizing constant, and $M$ is the number of samples in the output measurement vector, $\vect{q}$. From this expression, we see that the likelihood is a Gaussian function over the data $\vect{q}$ for a given set of parameters $\vect{a}$. However, for our inference problem we are interested in the likelihood as a function of the parameters $\vect{a}$ for a given set of data $\vect{q}$ (i.e. the data are fixed, and the parameters are varying)\footnote{This is the distinguishing factor between a probability density and a likelihood function: $p(A|B)$ is a probability density over $A$ for fixed $B$, or a likelihood function over $B$ for fixed $A$. The probability distribution must integrate to 1, while the likelihood function does not have this requirement.}.

The likelihood's dependence on $\vect{a}$ enters through the model prediction $\hat{\vect{q}}(\vect{a})$. Therefore, as a function of $\vect{a}$, the likelihood is only Gaussian if $\hat{\vect{q}}(\vect{a})$ depends linearly on $\vect{a}$. In our case, since the impulse response $h(\vect{a})$ depends nonlinearly on the parameters $\vect{a}$ (due to the nonlinear dependence of the Gaussian on $\vect{\tau}$ and $\vect{\sigma}$), the likelihood function is generally non-Gaussian in $\vect{a}$. This will be discussed further in the following section.

\subsubsection{The posterior}

We now combine the prior and likelihood using Bayes' theorem to obtain the posterior probability density over the parameters:
\begin{equation}
    p(\vect{a}|\vect{q},N) = \frac{p(\vect{q}|\vect{a},N)p(\vect{a}|N)}{p(\vect{q}|N)}
    \label{eq:bayes-theorem}
\end{equation}
where $p(\vect{q}|N)$ is the model evidence, which acts as a normalization constant for the posterior. The posterior pdf, $p(\vect{a}|\vect{q},N)$, describes our state of knowledge about the parameters after observing the data, combining both the information from the data (via the likelihood) and our prior knowledge (via the prior), weighted by their relative uncertainties. This is a mathematical representation of learning: we begin with some prior knowledge, are confronted by new evidence from data, and update our knowledge according to our relative trust in these two factors \cite{Jaynes2003}.

It is clear from eq. \eqref{eq:bayes-theorem} that if the prior and likelihood are Gaussian, the posterior will also be Gaussian (the product of two Gaussians is also a Gaussian). However, in general the posterior pdf does not have a closed-form expression due to the nonlinearity of the forward model. We therefore use an approximate inference framework to estimate the posterior by first finding its peak, and then estimating its width. 

The process of identifying the parameters that lie at the peak of the posterior pdf is called maximum a-posteriori (MAP) estimation:
\begin{equation}
    \vect{a}^* = \arg\max_{\vect{a}} p(\vect{a}|\vect{q},N).
\end{equation}
This is equivalent to minimizing the negative log of the unnormalized posterior, which is more convenient for numerical optimization:
\begin{equation}
    \vect{a}^* = \arg\min_{\vect{a}} \left[\mathcal{J}\right]
\end{equation}
where the cost function $\mathcal{J}$ is defined as:
\begin{equation}
\begin{aligned}
    \mathcal{J} &= -\log p(\vect{q}|\vect{a},N) - \log p(\vect{a}|N)\\
    &= \frac{1}{2}(\hat{\vect{q}}(\vect{a}) - \vect{q})^\top \precis{q} (\hat{\vect{q}}(\vect{a}) - \vect{q})\\
    &\quad + \frac{1}{2}(\vect{a} - \vect{a}_0)^\top \precis{a} (\vect{a} - \vect{a}_0) + K
\end{aligned}
\label{eq:cost}
\end{equation}
where $K$ denotes normalization terms that do not depend on $\vect{a}$. We now see the convenience of using Gaussian prior and likelihood functions: the negative log-posterior reduces to a sum of squared errors, weighted by the inverse covariances. We note, however, that the choice of Gaussian distributions is not driven by this convenience, but rather because it is the most natural choice for both the prior and the noise model, as discussed previously.

This minimization problem can be solved cheaply using gradient-based optimization, with the gradients of $\mathcal{J}$ given by:
\begin{equation}
    \nabla_{\vect{a}}\mathcal{J} = \mat{J_a}^\top \precis{q} (\hat{\vect{q}}(\vect{a}) - \vect{q}) + \precis{a} (\vect{a} - \vect{a}_0)
    \label{eq:cost-gradient}
\end{equation}
where $\mat{J_a} = \nabla_{\vect{a}} \hat{\vect{q}} \isreal^{M \times P}$ is the Jacobian of the forward model predictions with respect to the parameters. For a simple forward model such as our sum of Gaussians, these can be computed analytically or by using automatic differentiation. For more complex forward models, adjoint methods may be more appropriate, as they can be used to avoid computing this Jacobian directly \cite{Kontogiannis2025}.

Having found the peak of the posterior pdf, we now estimate its width using the Laplace approximation. We begin by performing a second order Taylor expansion of $\mathcal{J}$ around the MAP estimate $\vect{a}^*$, noting that $\mathrm{d} \mathcal{J}/\mathrm{d} \vect{a} = 0$ at $\vect{a}^*$:
\begin{equation}
    \mathcal{J}(\vect{a}) \approx \mathcal{J}(\vect{a}^*) + \frac{1}{2}(\vect{a} - \vect{a}^*)^\top \mat{H} (\vect{a} - \vect{a}^*)
    \label{eq:taylor-expansion}
\end{equation}
where $\mat{H} = \nabla_{\vect{a}}\nabla_{\vect{a}}\mathcal{J}|_{\vect{a}=\vect{a}^*} \isreal^{P \times P}$ is the Hessian of $\mathcal{J}$ evaluated at the MAP estimate. By inspection of eq. \eqref{eq:taylor-expansion} (and comparison to eq. \eqref{eq:cost}), we see that to second order, the posterior pdf is Gaussian around the MAP estimate with inverse-covariance:
\begin{equation}
    \precis{a^*} = \mat{H} \approx \mat{J_a}^\top \precis{q} \mat{J_a} + \precis{a}
    \label{eq:posterior-covariance}
\end{equation}
The accuracy of the Laplace approximation further from the MAP estimate depends on the degree of nonlinearity of the forward model, and the amount and quality of the data available. For uncorrelated noise, the first term reduces to $\sigma_q^{-2} \mat{J_a}^\top \mat{J_a}$, which becomes large (i) when the data is very precise (i.e., $\sigma_q$ is small), (ii) when the model is very sensitive to the parameters (i.e., $\mat{J_a}$ has large entries), or (iii) when there is a large amount of data (i.e., $M$ is large, therefore $\mat{J_a}$ has many rows). In any of these cases, the posterior becomes sharply peaked around the MAP estimate, such that the posterior probability falls to near-zero before the effect of the nonlinearity becomes significant. In Section \ref{sec:results} we will show with an LES case study that, even with moderate amounts of data, the true posterior is very close to Gaussian.

Finally, we note that we have made a further approximation in eq. \eqref{eq:posterior-covariance} by neglecting a term involving the second derivatives of the forward model predictions, which is: $\nabla_{\vect{a}}\mat{J_a}^\top \precis{q} (\hat{\vect{q}}(\vect{a^*}) - \vect{q})$. This term is small (i) when the model is close to linear in the neighbourhood of significant posterior probability mass (i.e., the second derivatives are small in this neighbourhood), or (ii) when the model fits the data well (i.e., $\hat{\vect{q}}(\vect{a}^*) - \vect{q} \approx 0$). If neither of these conditions are satisfied, which may be the case when the model is highly nonlinear and the data misfit is high (e.g. because the data is noisy), eq. \eqref{eq:posterior-covariance} will return a biased prediction of the posterior covariance. In this case, the second order term can be computed explicitly, approximated during the optimization procedure, or simply computed once using finite differences after the MAP estimate has been found.

We now have the required tools to estimate the parameters of the impulse response for a given model order $N$, and quantify our uncertainty in these estimates. To recap, we formulate the inverse convolution problem as a Bayesian parameter inference problem, defining (i) physically motivated prior densities over the parameters, and (ii) a Gaussian likelihood function based on the forward convolution model and a model for the measurement noise. We then find the MAP estimate of the parameters using gradient-based optimization, and estimate the posterior covariance using the Laplace approximation. At this point we have, for a given model order $N$, an improved estimate of the parameters $\vect{a}^*$ and their uncertainty $\covar{a^*}$. We now turn to the problem of selecting the most appropriate model order given the data.

\subsection{Bayesian model selection}
\label{sec:modelRanking}

Once we have found the posterior parameter pdfs for each model order of interest, we can rank the models according their probabilities, given the data:
\begin{equation}
    p(N|\vect{q}) \propto p(\vect{q}|N)p(N)
    \label{eq:model-posterior}
\end{equation}
where $p(N)$ is the prior probability we ascribe to each model order. We typically have no reason to prefer one model order over another \textit{a-priori}, so we assign equal prior probability to all models. The key quantity allowing us to discriminate between the models is therefore the evidence (or marginal likelihood) $p(\vect{q}|N)$, which was introduced in eq. \eqref{eq:bayes-theorem} as the normalization constant for the posterior parameter pdf. The model evidence describes the probability of observing the data under a given model order, integrating over all possible parameter values:
\begin{equation}
    p(\vect{q}|N) = \int p(\vect{q}|\vect{a},N)p(\vect{a}|N)\, d\vect{a}.   
    \label{eq:model-evidence}
\end{equation}
This integral is generally intractable for nonlinear models, but we can use the Laplace approximation to obtain a closed-form estimate. Substituting the Gaussian approximation for the unnormalized posterior into eq. \eqref{eq:model-evidence} and performing the integration yields:
\begin{equation}
    p(\vect{q}|N) \approx p(\vect{q}|\vect{a}^*,N) p(\vect{a}^*|N)(2\pi)^{P/2} |\covar{a^*}|^{1/2} 
    \label{eq:evidence-approx}  
\end{equation}
The first factor on the right-hand side of eq. \eqref{eq:evidence-approx} is the likelihood function evaluated at the MAP point. This measures how well the model fits the data at the MAP point, and is called the best-fit likelihood (BFL). A model that fits the data well is rewarded with a large best-fit likelihood. 

The remainder of the right-hand side, which is referred to as the Occam factor (OF), penalizes models that are more complex than necessary to fit the data. It is useful to expand this term further to understand how it achieves this. Substituting the Gaussian function for the prior and taking the logarithm, we obtain:
\begin{equation}
    \begin{aligned}
    \log{\mathrm{OF}} = &-\frac{1}{2}(\vect{a}^* - \vect{\mu_a})^\top \precis{a} (\vect{a}^* - \vect{\mu_a}) \\
    &- \frac{1}{2}\log\frac{|\covar{a}|}{|\covar{a^*}|}
    \end{aligned}
\end{equation}
=The first term penalizes models whose MAP parameters are far from the prior mean. If the priors encode physical expectations, as proposed in Section \ref{sec:prior}, this penalizes models that require parameters to deviate substantially from those expectations in order to fit the data. The second term involves the ratio of the prior and posterior covariance determinants. The determinant of the parameter covariance matrix measures the volume of the parameter space occupied by the distribution, so this ratio measures the factor by which the distribution contracts when the data are assimilated. A large contraction indicates that the model parameters must be finely tuned to fit the data, suggesting a fragile model whose predictions are highly sensitive to small perturbations in the parameters. This behaviour is penalized by the second term.

We gain a complementary insight by substituting eq.~\eqref{eq:posterior-covariance} into the second term and applying the matrix determinant lemma:
\begin{equation}
    \begin{aligned}
    -\frac{1}{2}\log{\frac{|\covar{a}|}{|\covar{a^*}|}} &= -\frac{1}{2}\log{\frac{|\covar{q} + \mat{J_a}\covar{a}\mat{J_a}^\top|}{|\covar{q}|}} \\
    \end{aligned}
\end{equation}
where $\mat{J_a}\covar{a}\mat{J_a}^\top$ is the covariance of the model predictions induced by the prior uncertainty in the parameters. The numerator therefore measures the volume of output space that the model can access given the prior, while the denominator measures the volume occupied by the data alone. This term therefore penalizes excessively flexible models whose prior uncertainty projects onto a large region of output space relative to the region occupied by the data. Such models are capable of explaining many different datasets, and are penalized for their lack of specificity.

\subsection{Uncertainty estimation}
\label{sec:NoiseEstimate}

As mentioned in Section \ref{sec:likelihood}, the likelihood function depends on the noise variance $\sigma_{q}^2$, which describes our expectations about the magnitude of the discrepancy $\varepsilon$ between the observed output data and the model predictions. In the context of inferring a flame impulse response model from LES data, we are rarely able to prescribe this variance \textit{a-priori}. We therefore estimate the noise variance by maximizing the marginal likelihood (MML) in eq.~\eqref{eq:evidence-approx} with respect to $\sigma_{q}$ \cite{MacKay1999}. Note that in eq.~\eqref{eq:evidence-approx} depends on $\sigma_{q}$ both explicitly through the best-fit likelihood, and implicitly through $\vect{a^*}$. This produces the noise estimate:
\begin{align}
    \sigma_{q} &= \frac{1}{M-\gamma}(\vect{q} - \hat{\vect{q}}(\vect{a}^*))^\top(\vect{q} - \hat{\vect{q}}(\vect{a}^*))
    \label{eq:noise-estimate} \\
    \gamma &= P - \text{Trace}\left(\covar{a^*}\precis{a}\right) \label{eq:gamma}
\end{align}
where $\gamma$ describes the number of parameters that are well-determined by the data, taking values between 0 and $P$. The optimal noise estimate is therefore the mean squared error (MSE) of the model fit, corrected for the number of parameters that have absorbed noise from the data. To see why this correction is necessary, consider the limiting case $\gamma \to M$, in which the model has one degree of freedom per data point and can therefore fit the data exactly. A standard MSE estimate would then predict zero noise, conflating an overparameterized model with noise-free data. The estimate in eq.~\eqref{eq:noise-estimate} diverges in this limit, which should be expected: we cannot infer $M$ effective parameters \emph{and} the noise level from $M$ data points. This distinction is therefore important, particularly when we attempt to infer model parameters from short signals. A full derivation of eq. \eqref{eq:noise-estimate} and further discussion of its properties can be found in Appendix \ref{app:noise-estimate-derivation}.

In principle, estimating $\sigma_q^2$ by MML requires nested optimization: an inner loop computes the MAP parameters $\vect{a}^*(\sigma_q)$ for a fixed noise level, and an outer loop updates $\sigma_q$ to maximize the marginal likelihood. In practice, we avoid this expensive nesting and instead update $\vect{a}$ and $\sigma_q$ within a single iterative procedure, converging them simultaneously. We have found this joint update to be robust and substantially cheaper for all cases considered. In early iterations, $\sigma_q$ tends to be overestimated, which downweights the data-misfit term and produces a smoother optimization landscape. As the iterate approaches the MAP point, $\sigma_q$ decreases toward its MML value, enabling an accurate final estimate of the posterior covariance.

\subsection{Summary}

The Bayesian inference framework described above is summarized as the following algorithm:
\begin{algorithm}[H]
\caption{Bayesian inference of flame impulse response}
\begin{algorithmic}[1]
\STATE Input data vectors $(\vect{u},\vect{q})$ 
\STATE Select range of model orders $N$ to consider
\FOR{each model order $N$}
    \WHILE{$\nabla_{\vect{a}} \mathcal{J} > \mathrm{tol}$}
        \STATE Use eq. \eqref{eq:cost-gradient} to step towards the MAP estimate
        \STATE Use eq. \eqref{eq:noise-estimate} to update noise estimate
    \ENDWHILE
    \STATE Use eq. \eqref{eq:posterior-covariance} to compute $\covar{a^*}$ 
    \STATE Use eq. \eqref{eq:evidence-approx} to estimate evidence $p(\vect{q}|N)$ 
\ENDFOR
\STATE Select model order $N$ with highest evidence
\end{algorithmic}
\end{algorithm}

\section{Robust solution of the Bayesian inverse problem}
\label{sec:robust}

In order to ensure success of the approximate inference framework described above, care must be taken in how the impulse response is parameterized and how the optimization problem is solved. We now describe our choices for these aspects of the problem.

\subsection{Parameterization and the prior}
\label{sec:parameterization}

A naive parameterization of the impulse response would be to use the raw parameters $\vect{a} = \{\vect{n},\vect{\tau},\vect{\sigma}\}$ directly. However, this has several drawbacks. First, the scales of the parameters can vary widely, leading to ill-conditioning in the optimization problem and poor convergence. A more robust parameterization would therefore normalize the scales of the parameters, keeping them all within a similar range.

The second drawback is that the time delays are symmetric under reordering, meaning that the parameter vector $\vect{\tau} = [\tau_1, \tau_2]$ describes the same impulse response as $\vect{\tau} = [\tau_2, \tau_1]$. This leads to a multi-modal posterior pdf, where each mode corresponds to a different ordering of the time delays. Our inference framework will find only one of these peaks, and ascribe all probability to the identified peak. The second requirement is therefore that we impose a specific ordering on the delays. 

The third drawback is that the time delays $\vect{\tau}$ and widths $\vect{\sigma}$ must be strictly positive to ensure causality and physical realism. Using the raw parameters would therefore require imposing hard constraints on the optimization problem, which can lead to numerical instability and convergence issues, and likely lead to a non-Gaussian posterior. We must therefore find a more robust parameterization that naturally enforces positivity.

To normalize the scales of the parameters, we exploit our knowledge that the characteristic timescales of the system are related to perturbations being convected along the flame. For a flame with length $L_f$, and bulk velocity $u$, we define the convective timescale $T_c = L_f/u$. We then normalize the time delays and widths by this timescale:
\begin{equation}
    \tilde{\tau}_i = \frac{\tau_i}{T_c}, \quad \tilde{\sigma}_i = \frac{\sigma_i}{T_c}
\end{equation}
This normalization ensures that all the parameters have similar orders of magnitude, improving the conditioning of the optimization problem. Note that if the characteristic timescale is difficult to define, which might be the case for some flames, $T_c$ can be treated as a hyperparameter that is inferred by maximizing the model evidence, as described for the noise variance in Section \ref{sec:NoiseEstimate}.

To remove the symmetry under reordering, we introduce a new set of parameters $\vect{\alpha}$, defined as the differences between consecutive time delays:
\begin{equation}
    \alpha_i = \tilde{\tau}_i - \tilde{\tau}_{i-1}, \quad i = 1, \ldots, N
\end{equation}
where we define $\tilde{\tau}_0 = 0$. If we also constrain $\alpha_i > 0$, then the time delays are automatically ordered as $\tau_1 < \tau_2 < \ldots < \tau_N$. This avoids symmetry under reordering, removing the symmetric peaks in the posterior pdf.

Finally, in order to enforce positivity of the parameters, we introduce a logarithmic transformation for both $\alpha_i$ and $\sigma_i$:
\begin{align}
    \gamma_i &= \log \alpha_i, \quad i = 1, \ldots, N \\
    \beta_i &= \log \tilde{\sigma}_i, \quad i = 1, \ldots, N
\end{align}
This transformation smoothly maps the positive real line to the entire real line, allowing us to define an unconstrained Gaussian probability density over ($\vect{\gamma},\vect{\beta}$) while keeping the physical parameters ($\vect{\tau},\vect{\sigma}$) strictly positive. 

Using these transformed parameters, we can define a broadly applicable Gaussian prior pdf that naturally enforces the physical constraints. While the prior can be adjusted if more specific knowledge about the parameters is available, a choice which should apply to a broad range of flames is:
\begin{align}
    p(n_i) &= \mathcal{N}(n_i;0, 1)\\
    p(\gamma_i) &= \mathcal{N}(\gamma_i;0, 0.5)\\
    p(\beta_i) &= \mathcal{N}(\beta_i;-1.8, 0.5)
\end{align}

\begin{figure*}[htbp]
    \centering
    \includegraphics[width=\textwidth]{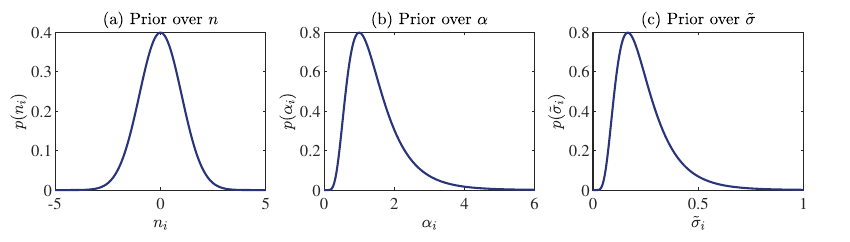}%
    \caption{Prior probability density functions for the impulse response parameters: (a) amplitude $n_i$, (b) non-dimensionalized time delay gap $\alpha_i$, (c) non-dimensionalized width $\tilde{\sigma}_i$.}
    \label{fig:priors}
\end{figure*}

To assist with interpretation, the resulting priors over the physical parameters are illustrated in Figure \ref{fig:priors}. We see that the prior over $n$ encodes an expectation that $n_i$ will lie between $\pm3$. This is motivated by several previous studies where $n$ has either been fixed to $\pm1$, or found to be close to $\pm1$ \cite{Schuermans2004,Blonde2023,Bade2013}, but provides additional range for flexibility. The Gaussian prior over $\gamma_i$ results in a log-normal prior over $\alpha_i$ (and equivalently $\tau_i$), which enforces positivity. The chosen width encodes an expectation that the time delays will be spaced between roughly 0.1 and 4 convective timescales, which is in line with observations from several previous studies. Similarly, the Gaussian prior over $\beta_i$ implies a log-normal prior over the non-dimensional width $\sigma_i/T_c$. The prior encodes an expectation that the dispersion time scale, $6\sigma_i$, is $\bigO(T_c)$. We therefore set the prior mean: $\tilde{\sigma_i} \approx 1/6 \approx \mathrm{exp}(-1.8)$. These priors therefore enforce broad but useful physical knowledge about the expected form and range of the parameters, while remaining flexible enough to adapt to a wide range of flames.

\subsection{Solution of the optimization problem}

Several gradient-based optimizers could be used to minimize $\mathcal{J}$ in equation \eqref{eq:cost}, given the gradients in equation \eqref{eq:cost-gradient}. We recognize that the cost function has the form of a nonlinear least-squares problem, which are most efficiently solved using Gauss-Newton-type optimizers, such as the Levenberg-Marquardt (LM) algorithm \cite{More1983,Nielsen1999}. This algorithm uses a cheap approximation of the second-order gradients of the cost function to achieve rapid convergence, which happens to be the same approximation as that given in eq. \eqref{eq:posterior-covariance}. 

Importantly, we note that gradient-based optimizers are local optimizers, and the nonlinearity of the forward model means that the cost function may have multiple local minima, even if the posterior is essentially unimodal. This is because we minimize the negative log-posterior, which may have local minima in regions that correspond to negligibly low posterior probability, i.e. there may be significant valleys in the cost landscape, that are not necessarily significant modes in the posterior pdf. To ensure we find the correct MAP point, we use a multi-start approach, where we run the optimizer from several initial conditions sampled randomly from the prior pdf. We then select the solution with the lowest cost as the MAP point, allowing us to robustly find the global minimum. In practice, the random starts required to reliably find the global minimum scales with the number of parameters. We propose 10 random starts per parameter, which works well on the problems considered in this work. The cost of further random starts is moderate, as they can be run in parallel.

\subsection{Edge effects and the valid convolution region}

In Section \ref{sec:introduction} we introduced the discrete convolution model in eq. \eqref{eq:discrete-conv}, which we repeat below for convenience:
\begin{equation*}
    q_m = \sum_{l=0}^{L-1} u_{m-l} h_l.
\end{equation*}
For a finite input record $\{u_m\}_{m=0}^{M-1}$ and an impulse response of length $L$, the first $L-1$ output samples involve input values prior to the start of the record (e.g.\ $u_{-1},u_{-2},\dots$). These inputs are unobserved, so the corresponding start-up samples cannot be modelled without introducing an additional assumption about the input prehistory. A common choice in system identification is to implicitly assume zero prehistory, which can bias the inferred impulse response when the record is short relative to $L$.

In this work, we avoid introducing a prehistory assumption by restricting inference to the \emph{valid} portion of the record for which the convolution is fully supported by measured inputs. Concretely, we discard the first $L-1$ output samples and evaluate the likelihood on the remaining $M_v = M-(L-1)$ samples. Equivalently, this corresponds to assigning the first $L-1$ output samples arbitrarily large uncertainty to reflect the impact of the unobserved input prehistory. A less conservative approach would be to reformulate the likelihood function to consider uncertainty in the \emph{input} signal as well as the output signal, and model the input prehistory as an additional uncertain variable with mean and uncertainty set to match the input signal statistics. This would allow us to use the full record, but in most cases the additional information gained is expected to be small, so we do not pursue this here.

\subsection{Choosing the impulse response support}
Choosing $T_h$ (or $L$) has two important consequences for the inference problem. First, it determines how much of the measured record can be used because we evaluate the likelihood only on the valid region, as discussed above. Increasing $T_h$ therefore reduces the amount of data from which the impulse response can be inferred.

Second, it controls the effective flexibility of the model. Allowing the impulse response to extend to larger delays gives the model more freedom to explain features in the output, and this increased flexibility should be penalized in model comparison through the Occam factor in the evidence.

We therefore treat the support in one of two ways. If the goal is parameter estimation for a fixed model order, $T_h$ can be prescribed from physical considerations (e.g. an expected maximum convective delay). This allows the available data to be used as efficiently as possible. If the goal is to rank model orders $N$ using the evidence, then $T_h$ must scale consistently across $N$ and consider all possible impulse responses of order $N$ that are supported by the parameter priors. For a given set of parameter priors, higher-order models admit longer plausible impulse responses. Holding $T_h$ fixed would therefore underestimate the flexibility of higher-order models, and bias the comparison towards them. For model ranking, we therefore set a support $T_h(N)$ using a conservative bound implied by our prior, i.e. $T_h(N)$ is chosen such that 99\% of impulse responses drawn from the prior for order $N$ lie within $[0,T_h(N)]$. This produces a fair comparison across $N$, but requires more data to reliably rank models.

\section{Results}
\label{sec:results}

We demonstrate the method on an LES dataset from the BRS burner, which is a laboratory-scale combustor with a turbulent, swirl-stabilized flame. The BRS burner has previously been used to demonstrate various methods for inferring flame impulse responses and transfer functions \cite{Komarek2010,Guo2019,Ghani2023,Eder2023a}. The rig is illustrated in Figure which shows that the swirler can be moved axially to study the effect on the flame response.

\begin{figure}
    \centering
    \includegraphics[width=\linewidth]{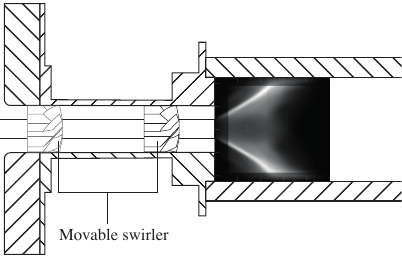}%
    \caption{Schematic of the BRS burner, showing the swirler in the forward position (black) and aft postion (grey).}
    \label{fig:BRS_burner}
\end{figure}

Komarek \& Polifke \cite{Komarek2010} performed compressible RANS simulations of the BRS burner, in which the inlet to the combustor was modelled as a boundary condition rather than resolving the swirler. In doing so, the authors could apply independent axial and circulatory fluctuations to the incoming flow and measure the flame response. By inspection, they found that the impulse response of the flame to velocity perturbations is well described by a sum of three Gaussians. The first has a positive peak and corresponds to the response of the flame to perturbations in axial velocity. The remaining two have one positive and one negative peak with gains that sum to zero, and correspond to the response of the flame to circulation fluctuations. This three-Gaussian model was later adopted by Ghani et al. \cite{Ghani2023} to demonstrate a method for inferring the parameters of the impulse response model from pressure time-series data.

\begin{figure}[ht]
    \centering
    \includegraphics[width=\linewidth]{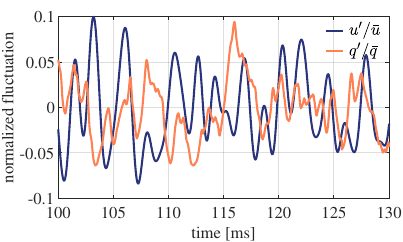}%
    \caption{Example input-output data from LES of a turbulent swirl flame: (blue) velocity perturbation, (orange) heat release rate fluctuation.}
    \label{fig:LES_data}
\end{figure}

In this study, we use data from Eder et al. \cite{Eder2023a}, who performed both compressible and incompressible LES of the BRS burner at a thermal power of 30~kW with the swirler in the forward position. The inlet was subject to bandwidth-limited broadband forcing, with a cut-off frequency of 630~Hz, and an amplitude of 10\% of the bulk velocity. The input-output signals were sampled at a frequency of 1~MHz for 0.5~s. The initial 0.03~s were discarded to remove simulation transients, and the input-output signals were normalized and down-sampled to 3.3~kHz. A sample of the input-output data is shown in Figure \ref{fig:LES_data}.

Eder et al. \cite{Eder2023a} applied system identification to the input-output data and found good agreement between the compressible and incompressible simulations, and with experimental data. They used built-in MATLAB function \texttt{impulseest} with cubic-spline regularization, signal pre-whitening, and a model order of 39. The appropriate down-sampling and model order need to be hand-tuned to achieve good results, and effectively constrain the length of the impulse response. In this case, a time increment of 0.3~ms and a model order of 39 constrains the impulse response to be 11.7~ms long.

We now apply the Bayesian inference framework described above to the same data, and compare the results with those obtained from system identification, and with experimental data. Unlike previous work, we do not assume a three-Gaussian model for the impulse response \emph{a priori}, but rather use the model selection framework to determine the most likely number of Gaussians given the data. We first assess a baseline case, where we replicate the signal properties used by Eder et al. \cite{Eder2023a}, and then explore the effect of reducing the signal length on both Bayesian inference (BI) and system identification (SI). 

\subsection{Baseline results:}

The required inputs for the proposed framework are (i) the input and output signals, (ii) the range of model orders to consider, and (iii) an estimate of the convective timescale $T_c$. As a baseline, we use the full 470~ms long signals, which we downsample to 3.3~kHz, as per Eder et al. \cite{Eder2023a}. We rank models with $N$=1-5 Gaussians, which bounds the three-Gaussian model that has previously been used for this burner. For the 30~kW operating condition studied, the inlet velocity was 11.3~m/s, which we use as the characteristic velocity scale. The flame length was taken to be the distance from the flame base to the point of impingement on the chamber wall, which was approximately 50~mm. This results in a convective timescale of $T_c=4.4$~ms. 

Like Eder et al., we find that the results from the compressible and incompressible simulations are practically identical, so we only report the incompressible case, as this is more appealing due to the lower computational cost of the simulations.

\begin{figure}[ht]
    \centering
    \includegraphics[width=\linewidth]{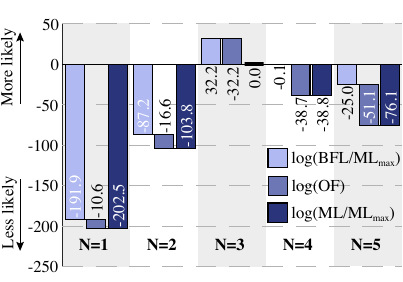}%
    \caption{Model ranking metrics for the baseline case, showing the three model ranking metrics described in Section \ref{sec:modelRanking}. The best-fit likelihood (BFL) rewards models that fit the data well, while the Occam factor (OF) penalizes models that are too flexible. The marginal likelihood (ML) balances these two effects, and is used to select the most likely model. The best-fit likelihood and marginal likelihood are shown relative to the marginal likelihood of the three-Gaussian model, which is the most likely model.}
    \label{fig:modelComparison_incomp_baseline}
\end{figure}

The three model ranking metrics described in Section \ref{sec:modelRanking} are shown for each of the five models in Figure \ref{fig:modelComparison_incomp_baseline}. Recall that the model with the highest marginal likelihood is the simplest model capable of describing the data, and should be selected as the most likely model. We see that this is the three-Gaussian model, which is consistent with the conclusion that Komarek \& Polifke arrived at by applying isolated perturbations at the flame base \cite{Komarek2010}. 

\begin{figure}[ht]
    \centering
    \includegraphics[width=\linewidth]{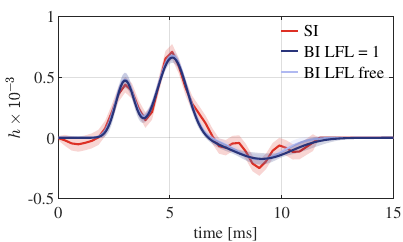}%
    \caption{Impulse responses inferred from LES data using system identification (SI) and Bayesian inference (BI) with and without enforcing the low-frequency limit (LFL). }
    \label{fig:h_incomp_baseline}
\end{figure}

The impulse responses inferred by both SI and the most likely model identified using BI with and without a constraint on the low-frequency limit (LFL) are shown in Figure \ref{fig:h_incomp_baseline}. We see that the SI and BI results are similar, but the BI impulse response contains fewer spurious undulations, which are not supported by our physical understanding of the system. These features change with the regularization and sampling parameters used for SI, indicating that they are artefacts of the method rather than robust features of the data. In the BI response, these artefacts are naturally filtered out by the prior information we provide (both in terms of the model structure and the parameter priors). Further, we see that enforcing the low-frequency limit has a negligible effect on the inferred impulse response.

The model parameters identified for this flame are summarized in Table \ref{tab:parameters_incomp_baseline}. Like Komarek \& Polifke \cite{Komarek2010}, we find one positive peak, and a positive-negative pair with gains that cancel, which is consistent with the flame response to axial and circulatory velocity perturbations, respectively. However, unlike Komarek \& Polifke, who studied a different operating condition with a thermal power of 70~kW, we find that the positive peak of the circulatory response arrives before the axial perturbation. This is likely due to changes in the relative convective velocities of the two types of perturbations with operating condition, but further analysis is required to fully understand this.

\begin{table}[ht]
\caption{Inferred impulse response parameters: amplitude $n_i$, nondimensionalized time delay $\tilde{\tau} = \tau/T_c$, and nondimensionalized width $\tilde{\sigma} = \sigma/T_c$. Parameters are shown for the baseline case without enforcing the low-frequency limit. Each parameter value is shown along with the standard deviation of the posterior distribution.}
\centering
\begin{tabular*}{\columnwidth}{@{\extracolsep{\fill}}cccc}
\hline
$i$ & $n_i$ & $\tilde{\tau}_i$ & $\tilde{\sigma}_i$ \\
\hline
1 & $ 0.5 \pm 0.03$ & $0.61 \pm 0.007$ & $0.09 \pm 0.008$ \\
2 & $ 1.1 \pm 0.04$ & $1.05 \pm 0.006$ & $0.14 \pm 0.007$ \\
3 & $-0.5 \pm 0.05$ & $1.89 \pm 0.026$ & $0.25 \pm 0.036$ \\
\hline
\end{tabular*}
\label{tab:parameters_incomp_baseline}
\end{table}

\begin{figure}[ht]
    \centering
    \includegraphics[width=\linewidth]{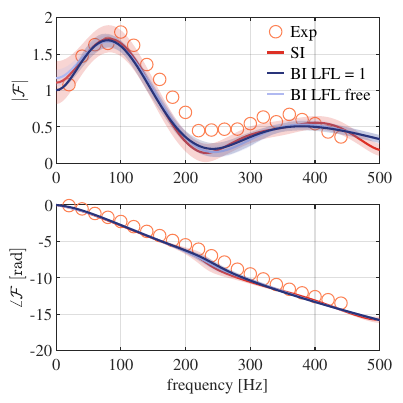}%
    \caption{Flame transfer functions obtained from the impulse responses shown in Figure \ref{fig:h_incomp_baseline}, compared with experimental data.}
    \label{fig:FTF_incomp_baseline}
\end{figure}

Next, we compare the resulting flame transfer functions (FTFs) in Figure \ref{fig:FTF_incomp_baseline}. We see that the FTFs obtained from both SI and BI are similar, indicating that the spurious peaks filtered out by BI do not significantly contribute to the FTF. We see that, without enforcing the low-frequency limit, the FTFs from both SI and BI take on a non-physical value slightly above 1, which is easily corrected in the BI framework, but not in the SI framework. Finally, we see that the FTFs from both methods match experimental data well. While this is more of a comment on the quality of the LES than the identification method, it adds confidence that the inferred impulse responses are physically meaningful.

\subsection{Effect of signal length}

Given the computational cost of reacting LES, there is a strong incentive to minimize the simulation time required to identify the flame response. We therefore assess how reducing the available data length affects both system identification and Bayesian inference. 

For this demonstration, we will enforce a fixed support of $T_h=3.5T_c$ for both SI and BI. As discussed in Section \ref{sec:robust}, applying a fixed support allows us to use the data more efficiently, but biases model comparison. We therefore assimilate the data into only the three-Gaussian model rather than performing model ranking like in the baseline case.

Starting from the baseline input-output data, we truncate the valid portion of the signals and retain only the first 20\%, 10\% and 5\%, corresponding to total simulation durations (including the time to flush out the effect of the input prehistory, but excluding the time to flush out simulation transients) of approximately 105~ms, 60~ms and 35~ms, respectively. We then apply SI and BI to these shorter records, downsampling to the same sampling frequency of 3.3~kHz and using the same range of model orders and convective timescale as in the baseline case. We retain the same broad priors over the parameters, as described in Section \ref{sec:parameterization}, for all cases.

\begin{figure}[ht]
    \centering
    \includegraphics[width=\linewidth]{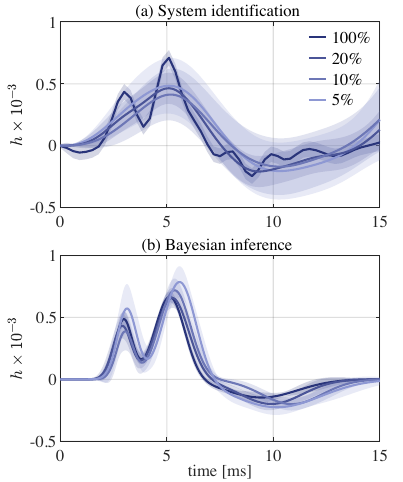}%
    \caption{Impulse responses inferred from truncated LES data using (a) system identification and (b) Bayesian inference with a fixed support of $T_h=3.5T_c$. We show results from the baseline case (100\%), and from cases where the valid portion of the data is truncated to 20\%, 10\% and 5\% of the original length.}
    \label{fig:h_incomp_short}
\end{figure}

\begin{figure*}[htb]
    \centering
    \includegraphics[width=\textwidth]{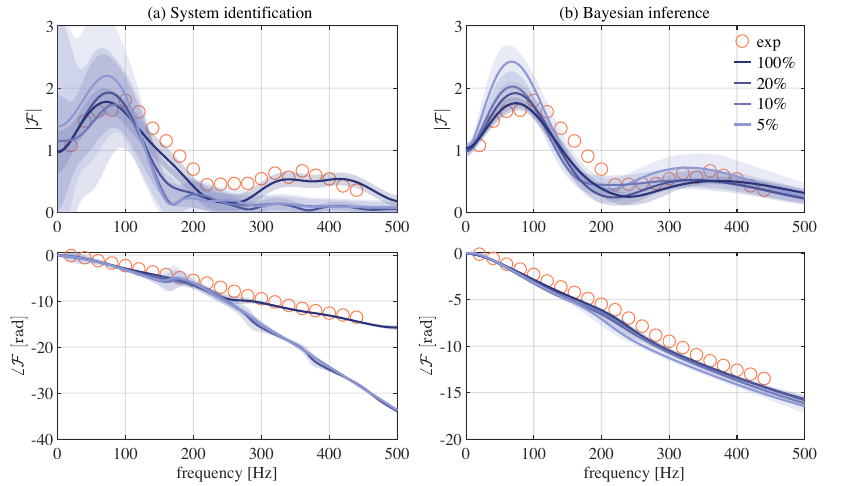}%
    \caption{Flame transfer functions obtained from the impulse responses shown in Figure \ref{fig:h_incomp_short}, compared with experimental data.}
    \label{fig:FTF_incomp_short}
\end{figure*}

The inferred impulse responses and the corresponding flame transfer functions are shown in Figures \ref{fig:h_incomp_short} and \ref{fig:FTF_incomp_short}. As the signal length decreases, both methods exhibit the expected loss of information, but with markedly different behaviour. To deal with the reduced information, the SI method must increase the regularization to avoid blowup. This leads to a substantial loss of temporal resolution in the impulse response, and a corresponding loss of fidelity in the FTF. In contrast, the BI method is able to draw from the prior information for regularization (again, this is both in the form of the model and in the parameter priors), which allows it to maintain a high temporal resolution in the impulse response, and a correspondingly high fidelity in the FTF, even as the data are shortened. 

\subsection{Assessing the Laplace approximation}

In section \ref{sec:bayesian}, we introduced the Laplace approximation, in which we approximate the posterior parameter distribution with a Gaussian centred at the MAP estimate. We demonstrated how the true posterior can be skewed by nonlinearities in the forward model, which is not captured under the Gaussian assumption. We claimed, however, that for moderate amounts of data, the posterior becomes sharply peaked around the MAP point, and most posterior mass lies in a neighbourhood where the model appears locally linear, meaning that the Laplace approximation is expected to be accurate. We now assess this claim by comparing the Laplace approximation with a sampling-based estimate of the posterior obtained using Markov Chain Monte Carlo (MCMC). MCMC makes no assumptions about the shape of the posterior, but is computationally expensive. In this case we use the Metropolis-Hastings algorithm \cite{Hastings1970}, which we run for 200,000 iterations, with a burn-in of 50,000 iterations (i.e. it requires 200,000 evaluations of the forward model). Our goal is to compare the posterior distributions obtained from MCMC with those obtained from the Laplace approximation, to assess the accuracy of the latter. We will do this for the three truncated signals considered above, to see how the accuracy of the Laplace approximation changes as the amount of data is reduced.

\begin{figure}[ht]
    \centering
    \includegraphics[width=\linewidth]{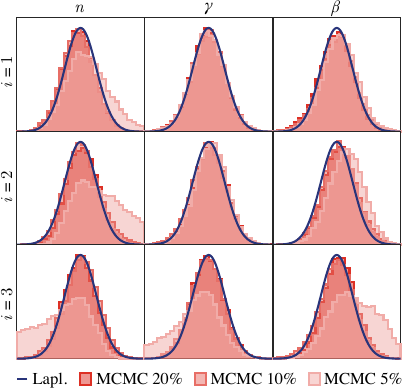}%
    \caption{Marginal posterior distributions of the model parameters obtained from MCMC sampling and the Laplace approximation, for the three truncated signals. Each marginal distribution is normalized to facilitate comparison of the shape of the posterior across data lengths.}
    \label{fig:MCMC_marginal}
\end{figure}

Figure \ref{fig:MCMC_marginal} compares the one-dimensional \emph{marginal} posterior distributions of the model parameters obtained from MCMC with those predicted by the Laplace approximation (here, ``marginal'' refers to the distribution of each parameter after integrating out all other parameters). For clarity, each marginal distribution is normalized so that we can compare the \emph{shape} of the posterior across the different data lengths.

For the shortest signal (5\% of the original valid record), the MCMC marginals exhibit noticeable skewness, and the Laplace approximation does not capture this asymmetry. Importantly, this discrepancy primarily affects the uncertainty quantification, while the MAP values remain well predicted. For the 10\% and 20\% signals, the marginal posteriors are already close to Gaussian, and the Laplace approximation tracks the MCMC results closely. This supports the claim that, once a moderate amount of information is available, the posterior concentrates around the MAP in a sufficiently small region for the second order Taylor expansion underpinning Laplace to be accurate.

\begin{figure}[ht]
    \centering
    \includegraphics[width=\linewidth]{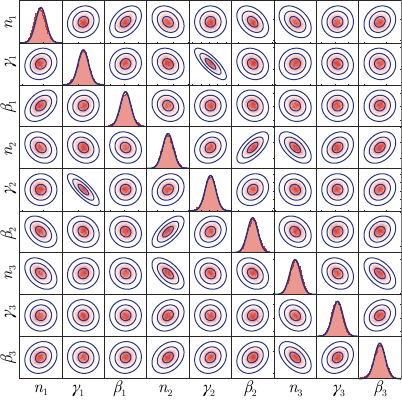}%
    \caption{Joint posterior distribution of the model parameters obtained from MCMC sampling (red) and the Laplace approximation (blue), for the 10\% truncated signal. Off-diagonal panels show pairwise joint distributions, while diagonal panels show the corresponding marginals. The joint distributions are shown as heatmaps for MCMC and as ellipses of 1-3 standard deviations for the Laplace approximation.}
    \label{fig:MCMC_joint}
\end{figure}

To provide a more complete view, Figure \ref{fig:MCMC_joint} shows the joint posterior for the 10\% case. Off-diagonal panels show pairwise joint distributions, while the diagonal panels reproduce the corresponding marginals. As above, the Laplace approximation provides an excellent match to the sampling-based posterior, and also reproduces the parameter correlations. The observed correlations are relatively weak in this case, suggesting that the parameters of the three-Gaussian model are uniquely identifiable from the available input-output data at this signal length.

\section{Conclusions}
\label{sec:conclusions}

This work reformulates the inverse problem of identifying a flame impulse response from recordings of velocity and heat release rate within a Bayesian framework. The impulse response is constrained to a physically motivated distributed time delay model represented as a sum of $N$ Gaussian pulses. For a given $N$, the parameters of the pulses are inferred from the data with quantified uncertainty bounds using Bayesian parameter inference. This is repeated for several values of $N$, and the most likely model order is selected using Bayesian model comparison.

This formulation replaces ad hoc choices in standard system identification, such as the model order, sampling rate, and regularization model and strength. The Bayesian framework instead provides a principled trade-off between data fit and model complexity, while producing impulse responses that are directly interpretable in terms of convective delays and dispersive broadening. 

On broadband-forced LES of the BRS swirl burner, this framework selected a three-Gaussian impulse response, which is consistent with prior physical interpretations of this configuration. Relative to standard system identification, the Bayesian approach produced an impulse response with fewer spurious features, while giving a comparable flame transfer function over the forced frequency range. It also enables straightforward enforcement of a known low-frequency limit, which is not easily achieved with standard system identification.

When the available record length is reduced, performance degrades mainly through increased posterior uncertainty rather than a qualitative change in the inferred impulse response. By contrast, standard system identification requires excessive regularization in the low-data regime, resulting in significant loss of temporal resolution in the impulse response.

Several extensions are natural. The present likelihood uses a white Gaussian noise model on the output only. Incorporating temporally correlated discrepancy and/or uncertainty in the measured input would broaden applicability. More generally, applying the method across operating conditions would allow the inferred delay parameters to be related more directly to underlying flame-flow physics and to support systematic, uncertainty-aware trends.

The software implementing the proposed framework is publicly available at \url{https://github.com/mattyoko/bayesian-flame-impulse-response/tree/paper-v1.0}. The repository includes scripts to reproduce the figures and analyses reported in this paper. We hope that this will facilitate application of the method to other flames and support systematic studies of how the inferred impulse response parameters relate to underlying flame-flow physics.

\section*{CrediT authorship contribution statement}
\noindent
\textbf{M.Y.}: Conceptualization, Formal analysis, Methodology, Software, Writing - original draft. 

\noindent
\textbf{W.P.}: Conceptualization, Writing - review \& editing.

\section*{Declaration of competing interest}

The authors declare that they have no known competing financial
interests or personal relationships that could have appeared to
influence the work reported in this paper.

\section*{Acknowledgments}

The authors thank Korbinian Niebler for preparing and transferring the LES data, and Alexander Eder who performed the simulations. This work was supported by the Leverhulm Trust [grant number ECF-2025-609].

\appendix
\section*{Appendix}
\section{Noise estimate derivation}
\label{app:noise-estimate-derivation}

We estimate the noise variance $\sigma_q^2$ by maximizing the log marginal likelihood of the model, given the data. In other words, we ask `what noise level makes the current model most likely, given this data?' This allows us to fairly rank models with different numbers of parameters. We now provide additional detail on this derivation, which is based on the work of MacKay \cite{MacKay1999}. For brevity, we write $\beta = \sigma_q^{-2}$ (the data precision) and work with $\beta$ throughout.

From eq.~\eqref{eq:evidence-approx} the Laplace-approximate log marginal likelihood is:
\begin{equation}
    \begin{aligned}
    \mathcal{L}(\beta) = \log p(\vect{q}|N) &\approx \log p(\vect{q}|\vect{a}^*,N) + \log p(\vect{a}^*|N) \\
    &+ \frac{P}{2}\log 2\pi + \frac{1}{2}\log|\covar{a^*}|
    \label{eq:app-logML}
    \end{aligned}
\end{equation}
Expanding the first term, which is the likelihood evaluated at the MAP point, we have:
\begin{equation}
    \log p(\vect{q}|\vect{a}^*,N) = \frac{M}{2}\log\frac{\beta}{2\pi} - \frac{\beta}{2}\|\vect{q}-\hat{\vect{q}}(\vect{a}^*)\|^2
    \label{eq:app-loglik}
\end{equation}
where $\|\vect{q}-\hat{\vect{q}}(\vect{a}^*)\|^2 = (\vect{q}-\hat{\vect{q}}(\vect{a}^*))^\top(\vect{q}-\hat{\vect{q}}(\vect{a}^*))$ is the sum of squared residuals.

The second term, which is the log prior evaluated at the MAP point, does not depend on $\beta$ explicitly, so it contributes only a constant. Substituting eq.~\eqref{eq:app-loglik} into eq.~\eqref{eq:app-logML} and dropping all terms that are constant with respect to $\beta$:
\begin{equation}
    \mathcal{L}(\beta) = \frac{M}{2}\log\beta - \frac{\beta}{2}\|\vect{q}-\hat{\vect{q}}(\vect{a}^*)\|^2 - \frac{1}{2}\log|\precis{a^*}|
    \label{eq:app-Lbeta}
\end{equation}

To maximize $\mathcal{L}(\beta)$ with respect to $\beta$, we set the derivative to zero. The first two terms are straightforward to differentiate. For the log-determinant term, we use the standard matrix identity:
\begin{equation}
    \frac{\mathrm{d}}{\mathrm{d}\beta}\log|\mat{A}| = \mathrm{Tr}\!\left(\mat{A}^{-1}\frac{\mathrm{d}\mat{A}}{\mathrm{d}\beta}\right)
    \label{eq:app-logdet-identity}
\end{equation}
Applying this to $\mat{A} = \precis{a^*} = \beta\,\mat{J_a}^\top\mat{J_a} + \precis{a}$ gives:
\begin{equation}
    \frac{\mathrm{d}}{\mathrm{d}\beta}\log|\precis{a^*}| = \mathrm{Tr}\!\left(\covar{a^*}\mat{J_a}^\top\mat{J_a}\right)
    \label{eq:app-dlogdet}
\end{equation}
Collecting all three contributions:
\begin{equation}
    \frac{\mathrm{d}\mathcal{L}}{\mathrm{d}\beta} = \frac{M}{2\beta} - \frac{1}{2}\|\vect{q}-\hat{\vect{q}}(\vect{a}^*)\|^2 - \frac{1}{2}\mathrm{Tr}\!\left(\covar{a^*}\mat{J_a}^\top\mat{J_a}\right)
    \label{eq:app-dLdbeta}
\end{equation}
Setting $\mathrm{d}\mathcal{L}/\mathrm{d}\beta = 0$ and multiplying through by $2\beta$:
\begin{equation}
    M - \beta\|\vect{q}-\hat{\vect{q}}(\vect{a}^*)\|^2 - \beta\,\mathrm{Tr}\!\left(\covar{a^*}\mat{J_a}^\top\mat{J_a}\right) = 0
    \label{eq:app-stationarity}
\end{equation}
Expanding the trace term:
\begin{equation}
    \begin{aligned}
    \mathrm{Tr}\!\left(\covar{a^*}\,\beta\mat{J_a}^\top\mat{J_a}\right) &= \mathrm{Tr}\!\left(\covar{a^*}\left[\precis{a^*} - \precis{a}\right]\right) \\
           &= \mathrm{Tr}\!\left(\mat{I}\right) - \mathrm{Tr}\!\left(\covar{a^*}\precis{a}\right) \\
           &= P - \mathrm{Tr}\!\left(\covar{a^*}\precis{a}\right) = \gamma
    \end{aligned}
    \label{eq:app-gamma-trace}
\end{equation}

To interpret $\gamma$, note that $\covar{a^*}\precis{a}$ has eigenvalues $\mu_j = (1+\beta\lambda_j)^{-1}$, where $\lambda_j$ are the eigenvalues of $\covar{a}\mat{J_a}^\top\mat{J_a}$, which measure the prior-weighted sensitivity of the model predictions to each parameter direction. Each eigenvalue $\mu_j$ lies between 0 and 1, and measures the fraction of the prior uncertainty in the $j$-th parameter direction that remains after observing the data. Therefore, $P - Tr(\covar{a^*}\precis{a}) = P - \sum_j(\mu_j)$ counts the number of parameter directions in which the data have substantially reduced the prior uncertainty, i.e.\ the number of parameters that are effectively determined by the data rather than by the prior. It follows that $0 \leq \gamma \leq P$.

Finally, we substitute eq.~\eqref{eq:app-gamma-trace} into eq.~\eqref{eq:app-stationarity} and solve for $\beta^{-1} = \sigma_q^2$:
\begin{equation}
    \sigma_q^2 = \frac{1}{M - \gamma}{(\vect{q}-\hat{\vect{q}}(\vect{a}^*))^\top(\vect{q}-\hat{\vect{q}}(\vect{a}^*))}
    \label{eq:app-sigma-final}
\end{equation}

\bibliographystyle{unsrt}
\bibliography{References}

\end{document}